\documentclass[12pt]{article}
\usepackage{amssymb,epsfig}

\renewcommand{\thefootnote}{\fnsymbol{footnote}}
\newcommand{\disp}{\displaystyle}
\newcommand{\es}{$\mathrm{E_6}$}

\begin{document}

\hspace*{\fill}MADPH-01-1237

\hspace*{\fill}WUE-ITP-2001-013

\hspace*{\fill}hep-ph/0107080

\vspace{1cm}

\begin{center}
\textbf{\Large Neutralinos in E$_\mathbf{6}$ inspired supersymmetric\\
U$\mathbf{(1)'}$ models}

\vspace{10mm}
S.~Hesselbach\footnote{e-mail: hesselb@pheno.physics.wisc.edu}

Department of Physics, University of Wisconsin, Madison,
WI 53706, USA

\vspace{5mm}
F.~Franke\footnote{e-mail: fabian@physik.uni-wuerzburg.de},
H.~Fraas\footnote{e-mail: fraas@physik.uni-wuerzburg.de}

Institut f\"ur Theoretische Physik und Astrophysik,
Universit\"at W\"urzburg,\\
D-97074 W\"urzburg, Germany
\end{center}

\vspace{5mm}
\begin{abstract}
The neutralino sector in \es\ inspired supersymmetric models with extra 
neutral gauge bosons and singlet Higgs fields contains additional gaugino 
and singlino states compared to the MSSM. We discuss the neutralino
mixing in rank 5 and rank 6 models and analyze the supersymmetric
parameter space where the light neutralinos have mainly singlino or
MSSM character. The neutralino character, 
resonance effects of the new gauge bosons 
and, assuming mSUGRA-type RGEs,
different selectron masses
lead to significant differences
between the MSSM and the extended models in neutralino production at an
$e^+e^-$ linear collider. Beam polarization may improve the signatures to
distinguish between the models. In an appendix, we present the mass terms
of the gauge bosons, charginos and sfermions which show a significant
different mass spectrum than in the MSSM 
and give all relevant neutralino couplings.
\end{abstract}

\renewcommand{\thefootnote}{\arabic{footnote}}
\setcounter{footnote}{0}

\section{Introduction}

Supersymmetry is considered to be one of 
the most fascinating concepts in particle
physics. It may provide
solutions to the hierarchy and fine tuning problems of the
Standard Model (SM) and
allows unification of the coupling constants at the scale 
$E_\mathrm{GUT}$ of a Grand Unified Theory (GUT).
In the Minimal Supersymmetric Standard Model (MSSM) \cite{haberkane} 
one has to explain the weak scale value of the $\mu$ parameter in
the superpotential \cite{giudice}.
Supersymmetric models with additional singlet Higgs fields 
evade the $\mu$ problem 
by replacing $\mu$ with a product of a dimensionless coupling and a vacuum
expectation value of a singlet Higgs field.
The simplest extension of the MSSM
by one additional singlet Higgs field 
is the Next-to-minimal
Supersymmetric Standard Model (NMSSM) \cite{nmssmsammel}.

Models with additional U(1) factors in the gauge group containing new
neutral gauge bosons are a further extension of the MSSM
\cite{cveticdemir, cveticlangacker, espinosadw}. These models provide
a solution to the domain-wall problem of the NMSSM
\cite{abelsarkar}, because the discrete $\mathbb{Z}_3$ symmetry is
embedded in the new U(1) factors \cite{cveticdemir,
espinosadw}. One or two additional U(1) factors can be motivated by
the breaking of an \es\ group which is a good candidate of an
(effective) GUT group \cite{hr}. 
Since experimental results lead to strict lower mass bounds for the
new gauge bosons, 
further studies are needed in order to understand
the hierarchy between the gauge boson masses \cite{ZsHierarchy}.
Also the SUSY breaking scale is typically of the order of the new gauge bosons.
However, gaugino and slepton masses of some 100 GeV at the electroweak scale
are not excluded \cite{drees, everett}.
The extended neutralino sector in
these \es\ inspired U$(1)'$ models 
will be discussed in detail in this work focusing especially on the
differences to the MSSM.

The production of neutralinos at an electron-positron linear collider
with polarized beams is considered as an excellent process to
discriminate between the supersymmetric models.
The experimental signatures certainly depend on the neutralino decay 
channels which are discussed for singlino-like lightest supersymmetric
particles (LSP) in \cite{hugonie}. Especially one expects the existence
of displaced decay vertices for large singlet vacuum expectation values
\cite{displacedpaper}.

We discuss the neutralino phenomenology in two types of
scenarios where the mass of the new gauge bosons is above the
reach of the first generation of linear colliders with 500 to 800~GeV
center-of-mass energy. Within the framework of constrained
\es\ models with GUT relations between the soft
supersymmetry breaking masses of order 100~GeV
the lighter
four neutralinos have MSSM-like character.
Assuming mSUGRA-type RGE relations for the
sfermion masses, 
the masses of the left and
right selectrons in the constrained \es\ models 
considerably differ from the ones in the constrained MSSM. 
Especially if the selectrons cannot be directly produced
at a linear collider, 
neutralino production may offer valuable information about
the underlying supersymmetric model.
The selectron RGE relations in \es\ models can be tested
in scenarios with gaugino-like light neutralinos
while particularly the polarization asymmetries of the 
production cross sections of higgsino dominated neutralinos
show new gauge boson resonance effects
at energies well below their masses.

In a second type of scenarios we relax the GUT and RGE relations and
obtain singlino-like light neutralinos for large values of 
the soft breaking U$(1)'$ gaugino mass parameter.
Then neutralino production 
provides a favorable way to determine the neutralino character and 
the parameters of the underlying supersymmetric model.

This paper is organized as follows: In section 2 we
analyze the neutralino sector of the \es\ inspired models with new
U(1) factors
and discuss the mass spectra of the
neutralinos in these models which contain up to four exotic singlino
and new gaugino states.
In section 3 the production of neutralinos in electron-positron
collisions is analyzed
in representative scenarios where the lightest MSSM-like neutralino
has gaugino or higgsino character.
We work out differences of the cross
sections and polarization asymmetries between the MSSM and the extended
models for MSSM-like light neutralinos and discuss as well the
production of exotic singlino-like neutralinos.
In the appendix we give a brief overview of the \es\ models focusing
mainly on the breaking of the \es\ group resulting in new U(1) factors
and on the particle content. We discuss
the mass terms of the gauge bosons, charginos and sfermions,
which have a significantly different mass spectrum than in the MSSM
and the NMSSM because of additional D-terms of the new U(1) factors.
The relevant couplings for production and decay of the
neutralinos in the \es\ models can also be found in the appendix.

\section{Neutralino mass spectra}

The additional gauge bosons and singlet Higgs fields in
\es\ inspired models lead to an extended neutralino sector which may be
crucial in order to distinguish between these models and the MSSM or NMSSM
in future high energy collider experiments. 
The breaking of a GUT group \es\ can lead to low energy
gauge groups with one (rank 5) or two (rank 6) additional
U(1) factors in comparison to the SM (App.~\ref{e6breaking}).
The particle spectrum of the \es\ models contains two neutral SM
singlet fields which can be interpreted as singlet Higgs fields
(App.~\ref{partcont}). The vacuum expectation values (vevs) of these
singlets break the new U(1) factors and create masses for
the corresponding new gauge bosons large enough to respect the
experimental bounds (App.~\ref{gaugesector}).
We assume the absence of gauge kinetic mixing between the U(1) factors
in the discussed models.

In the \emph{rank-5 model with one singlet} (R5$_1$) only one of the
singlet fields obtains a vev, whereas 
in the \emph{rank-5 model with two singlets} (R5$_2$) both singlets
vevs are present.
In the \emph{rank-6 model} (R6) both vevs
are necessary to create the masses for both new gauge bosons.
The neutralino masses and mixings
in these models summarized in Table \ref{models} depend on the
soft symmetry breaking gaugino mass parameters, the ratio
$\tan\beta = v_2/v_1$ of the doublet Higgs vevs, the singlet Higgs
vevs and the trilinear coupling $\lambda$ of the superpotential term
$W_\lambda = \lambda H_1 H_2 N_1$ which replaces the
$\mu$ term of the MSSM. 
$W_\lambda$ is the only superpotential term relevant for the mass
terms of the neutralinos. Terms $\sim\!\! H_1 H_2 N_2$ or $\sim\!\!
N_1^\dagger N_2$, where $H_1$, $H_2$, $N_1$, $N_2$ are the doublet 
and singlet Higgs fields
defined in App.~\ref{partcont},
are forbidden by the \es\ gauge 
symmetry \cite{hr, boyce, gunion, habersher}.
Thus the product of the dimensionless coupling
$\lambda$ with the singlet vev $v_3$ becomes the effective $\mu$ parameter
in the \es\ models.

\begin{table}[ht]
\begin{center}
\begin{tabular}{||c||c|c|c|c|c|c||}
\hline \hline
Model & Rank & new gauge & singlet  & neutralinos &
soft breaking  & singlet \\
& & factors & Higgs &  & parameters & vevs \\
\hline \hline
R5$_1$ & 5 & U(1)$'$ & 1 & 6 & $M_2$, $M_1$, $M'$ & $v_3$
\\ \hline
R5$_2$ & 5 & U(1)$'$ & 2 & 7 & $M_2$, $M_1$, $M'$ & $v_3$, $v_4$
\\ \hline
R6     & 6 & $\mathrm{U(1)'\times U(1)''}$ & 2 & 8 
 & $M_2$, $M_1$, $M'$, $M''$ & $v_3$, $v_4$
\\ \hline \hline
\end{tabular}
\end{center}
\caption{Rank of the gauge group, new factors in the gauge group,
number of the singlet Higgs fields obtaining a vev, resulting number of the
neutralinos, soft breaking gaugino mass parameters and singlet vevs
in the considered models R5$_1$ (rank-5 model with one singlet), R5$_2$
(rank-5 model with two singlets) and R6 (rank-6 model).}
\label{models}
\end{table}

The neutralino masses and mixings can 
be derived from the most general neutralino mass term
in the Lagrangian of the rank-6 model
\begin{eqnarray}
\mathcal{L}_{m_{\chi^0}} & = &\frac{1}{\sqrt{2}}ig_2\lambda^3(v_1\psi_{H_1}^1-
 v_2\psi_{H_2}^2)-\frac{1}{\sqrt{2}}ig_1\lambda_1
  (v_1 \psi_{H_1}^1-v_2 \psi_{H_2}^2) \nonumber \\[1mm]
  & & {}+ \frac{1}{\sqrt{2}}ig'\lambda ' \left( Y'_1 v_1\psi_{H_1}^1+
   Y'_2 v_2\psi_{H_2}^2 +Y'_3 v_3 \psi_{N_1} +Y'_4 v_4 \psi_{N_2}\right)
    \nonumber \\[1mm]
    & & {}+ \frac{1}{\sqrt{2}}ig''\lambda '' \left( Y''_1 v_1\psi_{H_1}^1+
     Y''_2 v_2\psi_{H_2}^2 +Y''_3 v_3 \psi_{N_1} +Y''_4 v_4 \psi_{N_2}\right)
      \nonumber \\[1mm]
      & & {}+ \frac{1}{2} M_2 \lambda^3 \lambda^3 + \frac{1}{2} M_1 \lambda_1
       \lambda_1 + \frac{1}{2} M' \lambda' \lambda'
	+ \frac{1}{2} M'' \lambda'' \lambda'' \nonumber \\[2mm]
	& & {}- \lambda v_3 \psi_{H_1}^1 \psi_{H_2}^2
	 -\lambda v_1 \psi_{H_2}^2 \psi_{N_1}
	  -\lambda v_2 \psi_{H_1}^1 \psi_{N_1} \nonumber \\[1mm]
	  & & {} + \mbox{h.c.} \;\; ,
\end{eqnarray}
with the two component Weyl spinors
$\lambda^3\!$, $\lambda_1$, $\lambda'$ and $\lambda''$ of the neutral
\mbox{SU(2)$_\mathrm{L}$-,} U(1)$_Y$-, U(1)$'$ and U(1)$''$ gauginos 
and $\psi_{H_1}^1$, $\psi_{H_2}^2$, $\psi_{N_1}$ and $\psi_{N_2}$
of the doublet and singlet higgsinos (singlinos), respectively.
The $Y'_i$ ($Y''_i$) are the U(1)$'$ (U(1)$''$) quantum numbers of the
doublet and singlet Higgs fields.
In the following we will assume $\mathrm{U}(1)' \equiv
\mathrm{U}(1)_\eta$ in the rank-5 models, so in R5$_1$ and R5$_2$
\begin{equation} \label{Ysr5}
  Y'_{1,2} = Y_\eta(H_{1,2}) \,, \qquad Y'_{3,4} = Y_\eta(N_{1,2}) \,,
\end{equation}
as given in table \ref{quantnr}. In the rank-6 model (R6) it is
$\mathrm{U}(1)' \equiv \mathrm{U}(1)_\psi$ and $\mathrm{U}(1)'' \equiv
\mathrm{U}(1)_\chi$, so with table \ref{quantnr}
\begin{equation} \label{Ysr6}
  Y^{'('')}_{1,2} = Y_{\psi(\chi)}(H_{1,2}) \,, 
  \qquad Y^{'('')}_{3,4} = Y_{\psi(\chi)}(N_{1,2}) \,.
\end{equation}
With the assumption that all U(1) factors are created at the same
energy scale, e.g.\ at the scale where an underlying \es\ group is
broken, and obey the same renormalization group equations the
couplings $g_1$, $g'$ and $g''$ should have the same value at the electroweak
scale \cite{hr,gunion,binetruy}. In the remainder of this paper we
assume $g' = g'' = g_1$ in all numerical discussions.

Then in the basis
\begin{equation} \label{r6basis}
  (\psi^0)^T=(-i\lambda_{\gamma},-i\lambda_Z,\psi_H^a,\psi_H^b,
              -i\lambda',\psi_{N_1}[, \psi_{N_2}[, -i\lambda'']])
\end{equation}
with
\begin{eqnarray}
  \lambda_{\gamma} & = & \lambda^3 \sin\theta_W 
   + \lambda_1 \cos\theta_W
   \, , \nonumber \\
  \lambda_Z & = & \lambda^3 \cos\theta_W 
   - \lambda_1 \sin\theta_W
   \, , \nonumber \\[-3mm] \label{neubasis} \\[-3mm]
  \psi_H^a & = &
  \psi_{H_1}^1 \cos \beta - \psi_{H_2}^2 \sin\beta \, , \nonumber \\
  \psi_H^b & = & \psi_{H_1}^1 \sin\beta + \psi_{H_2}^2 \cos\beta \, ,
  \nonumber
\end{eqnarray}
the neutralino mass term in the rank-6 model becomes
${\cal L}_{m_{\chi^0}} = -\frac{1}{2} (\psi^0)^T Y \psi^0 +
\mbox{h.c.}$ with
the $8\times 8$ neutralino mass matrix
\begin{equation} \addtolength{\arraycolsep}{3pt} \label{Ynr6}
  \renewcommand{\arraystretch}{2} 
  Y = \left( \begin{array}{cccccc|c|c}
      Y_{11} & Y_{12} & 0 & 0 & 0 & 0 & 0 & 0\\
      Y_{12} & Y_{22} & m_{Z^\mathrm{SM}} & 0 & 0 & 0 & 0 & 0\\
      0 & m_{Z^\mathrm{SM}} & -\lambda v_3 \sin2\beta & 
       \lambda v_3 \cos2\beta & Y_{35} & 0 & 0 & Y_{38}\\
      0 & 0 & \lambda v_3 \cos2\beta & \lambda v_3 \sin2\beta &
      Y_{45} & \lambda v & 0 & Y_{48}\\
      0 & 0 & Y_{35} & Y_{45} & M' & \disp Y'_3 \frac{g' v_3}{\sqrt{2}} & 
      \disp Y'_4 \frac{g' v_4}{\sqrt{2}} & 0\\
      0 & 0 & 0 & \lambda v & 
       \rule[-5mm]{0mm}{10mm} \disp Y'_3 \frac{g' v_3}{\sqrt{2}} & 
       0 & 0 &
       \disp Y''_3 \frac{g'' v_3}{\sqrt{2}}\\ \cline{1-6}
      0 & 0 & 0 & 0 & 
       \rule[-5mm]{0mm}{10mm} \disp Y'_4 \frac{g' v_4}{\sqrt{2}} &
       \multicolumn{1}{c}{0} & 0 &
       \disp Y''_4 \frac{g'' v_4}{\sqrt{2}}\\ \cline{1-7}
      0 & 0 & Y_{38} & Y_{48} & 0 & 
       \multicolumn{1}{c}{\disp Y''_3 \frac{g'' v_3}{\sqrt{2}}} &
       \multicolumn{1}{c}{\disp Y''_4 \frac{g'' v_4}{\sqrt{2}}} & M''
    \end{array} \right)
\end{equation}
where the matrix elements are given by
\begin{equation} \setlength{\arraycolsep}{3ex}
  \renewcommand{\arraystretch}{2}
  \begin{array}{@{}ll@{}}
    Y_{11} = M_2 \sin^2\theta_W  + M_1 \cos^2\theta_W \, , &
    \disp Y_{45} = \frac{g' v}{2\sqrt{2}} (Y'_1 + Y'_2) \sin2\beta \; , \\
    Y_{12} = (M_2-M_1) \sin\theta_W \cos\theta_W \, , &
    \disp Y_{38} = \frac{g''v}{\sqrt{2}} \left( Y''_1 \cos^2\beta - Y''_2
      \sin^2\beta \right) , \\
    Y_{22} = M_2 \cos^2\theta_W + M_1 \sin^2\theta_W \, ,  &
    \disp Y_{48} = \frac{g'' v}{2\sqrt{2}} (Y''_1 + Y''_2) \sin2\beta
    \; , \\
   \disp Y_{35} = \frac{g'v}{\sqrt{2}} \left( Y'_1 \cos^2\beta - Y'_2
      \sin^2\beta \right) , &
   \disp v \equiv \sqrt{v_1^2 + v_2^2} = \sqrt{2} \: \frac{m_W}{g_2} \; .
  \end{array}
\end{equation}
Assuming $CP$ conservation, all parameters are real.
The physical masses of the neutralinos can be derived by
diagonalization with a real orthogonal $8\times 8$ matrix $N$
\cite{haberkane, bartlfraasneut}
\begin{equation}
  \eta_{\tilde{\chi}^0_i}
  m_{\tilde{\chi}^0_i} \delta_{ik} = N_{im} N_{kn} Y_{mn} \, ,
\end{equation}
with the physical masses $m_{\tilde{\chi}^0_i}$ of the neutralinos and
the sign factors $\eta_{\tilde{\chi}^0_i}$ of the respective eigenvalues.
The $6 \times 6$ and $7 \times 7$ neutralino mixing matrices of the
models R5$_1$ and R5$_2$, respectively, are obtained as the upper
left $6 \times 6$ and $7 \times 7$ submatrices of $Y$ as shown in
eq.~(\ref{Ynr6}) \cite{cveticdemir, gunion, ellis, nandi, keithma,
  decarlosespinosa, suematsu, suematsu3, gherghetta}.

\begin{figure}[p]
\centering
\begin{picture}(16,17.9)
  \put(0,-.05){\epsfig{file=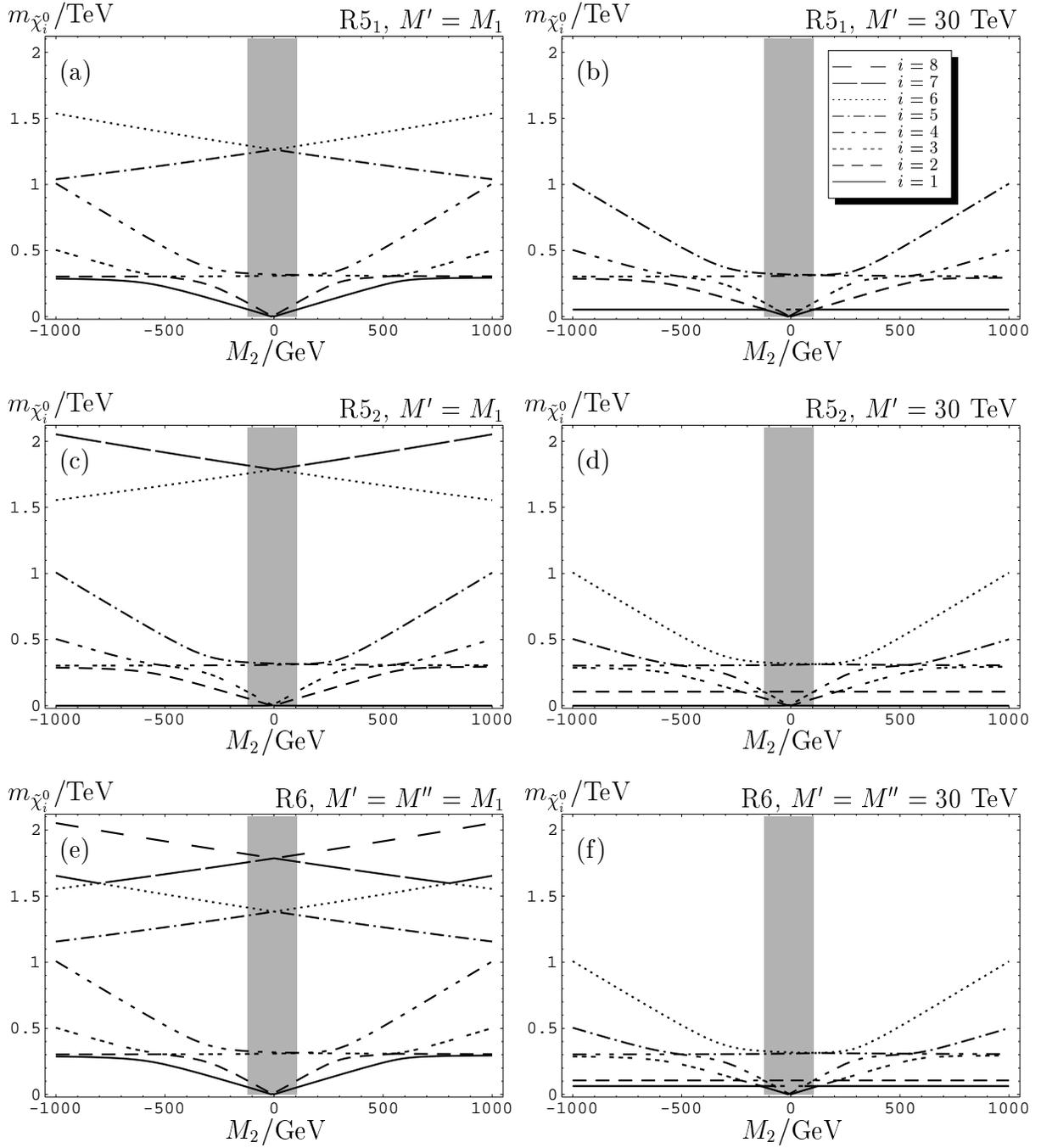}}
\end{picture}
\caption{\label{mspectra}Mass spectra of the neutralinos in the \es\
models for $\tan\beta = 5$, $\lambda=0.1$, $v_3 \:[= v_4] = 3$~TeV and
$M_1 = M_2\: 5/3 \tan^2\theta_W$:
(a) model R5$_1$ with $M'=M_1$,
(b) model R5$_1$ with $M'=30$~TeV,
(c) model R5$_2$ with $M'=M_1$,
(d) model R5$_2$ with $M'=30$~TeV,
(e) model R6 with $M'=M''=M_1$ and
(f) model R6 with $M'=M''=30$~TeV.
The shaded area marks the experimentally excluded parameter space
\cite{lepbounds}.}
\end{figure}

The upper left $4\times 4$ submatrix contains the mixing matrix of the MSSM
if $-\lambda v_3$ is replaced by the $\mu$ parameter.
The lower right submatrix ($2\times 2$, $3\times 3$ and $4\times 4$,
respectively) represents the new exotic components of the neutralinos
in the \es\ models.
The entries in the nondiagonal submatrices are of the order of the
doublet vacuum expectation values and therefore of
$m_{Z^\mathrm{SM}}$. On the other hand in the ``exotic'' submatrix
most entries are of the order of the singlet vacuum expectation values
and therefore of $m_{Z'}$ and $m_{Z''}$. This results in an
approximate decoupling of
the exotic neutralinos from the MSSM-like ones as shown in
Figs.~\ref{mspectra} (a) and (e) for the models R5$_1$ and R6,
respectively, with GUT relation $M' = M'' =
M_1 = M_2\: 5/3 \tan^2\theta_W$ for the gaugino mass parameters.
With $M_2,
\lambda v_3 = \mathcal{O}(100~\mathrm{GeV})$,
the four lighter neutralinos are mainly MSSM-like while the
heavy neutralinos have exotic character
with masses of the order of $m_{Z'}$ and $m_{Z''}$.
In model R5$_1$ (Fig.~\ref{mspectra} (a))
the masses of the two exotic neutralinos are approximately
\begin{equation} \label{mnexotic}
 m_{\tilde{\chi}^0_{5,6}} \approx Y'_3 \frac{g' v_3}{\sqrt{2}} \pm
  \frac{1}{2} M'
  = Y'_3 \frac{g' v_3}{\sqrt{2}} \pm M_2 \frac{5}{6} \tan^2 \theta_W \,.
\end{equation}
Eq.~(\ref{mnexotic}) is valid for all small $M' \ll
v_3$, including $M' < M_1$. Hence both corresponding exotic neutralinos
always have masses of order $m_{Z'}$ and are mixtures of 
the singlino and $Z'$ gaugino eigenstates.
For $M_1, M_2, \lambda v_3 = \mathcal{O}(v_3,v_4)$
the exotic neutralinos may be the lightest neutralinos, but have
nevertheless masses of the order $m_{Z'}$ even for small $M'$.

In model R5$_2$
the exotic $3\times 3$ submatrix is singular so
the lightest neutralino is very light ($m_{\tilde{\chi}^0_1}=0.2$~GeV)
and has mainly singlino character
(Fig.~\ref{mspectra} (c)). Nevertheless the exotic neutralinos
decouple from the MSSM-like ones in good approximation \cite{lc97}.

If the GUT relation for $M'$ is relaxed in model R5$_1$
and a large value $M' \gg v_3$ chosen, a LSP with singlino
character is possible \cite{decarlosespinosa}. Then a mechanism like
the see-saw effect in the neutrino mass matrix
\cite{seesaw} occurs in the submatrix
of the exotic neutralinos which results in a light singlino-like and
a very heavy $\tilde{Z}'$-gaugino-like neutralino (the
$\tilde{\chi}^0_6$) with masses
\begin{equation}
  m_\textrm{\scriptsize singlino-like} \approx {Y'_3}^2 {g'}^2
  \frac{v_3^2}{2 M'} \,,
  \quad  m_{\tilde{\chi}^0_6} \approx  M' \,.
\end{equation}
Fig.~\ref{mspectra} (b) shows the neutralino mass spectrum for $M' =
30$~TeV resulting in a singlino-like neutralino with mass $54$~GeV, which is
the LSP in a large fraction of the parameter space.

In model R5$_2$ a large value of $M' = 30$~TeV
leads to a second light singlino-like neutralino with mass $106$~GeV
(Fig.~\ref{mspectra} (d)).
In model R6 a large value for $M'$ and GUT relation $M'' =
M_1$ and vice versa results again in one light singlino, whereas for
large values of both $M'$ and $M''$ two light singlinos are possible
due to this see-saw effect in the $4\times 4$ submatrix of the exotic
neutralinos.
This is shown in Fig.~\ref{mspectra} (f) for $M' = M'' = 30$~TeV with
singlino-like neutralinos of masses 64~GeV and 106~GeV which are
the lightest and second lightest neutralino in large parameter regions.

To conclude because of the structure of the neutralino mixing matrix
in the considered \es\ inspired models the lighter four
neutralinos have MSSM-like character in models R5$_1$  and R6
with $M_1, M_2, M', M'',
\lambda v_3 = \mathcal{O}(100~\mathrm{GeV})$. Light
singlino-like neutralinos exist in the model R5$_2$ and the models
R5$_2$ and R6 with large $M'$, $M''$, whereas light
neutralinos never have dominant $Z'$ gaugino
character.

\section{Neutralino production at an electron-positron collider}

\subsection{Cross sections}

The production of neutralinos
$e^+ e^- \to \tilde{\chi}^0_i \tilde{\chi}^0_j$
in the \es\
models proceeds via $s$ channel exchange of the neutral gauge bosons
$Z_n$ and $t$ and $u$ channel exchange of selectrons.
The production cross section
\begin{equation} \label{twq}
\sigma = \left( \sigma_Z + \sigma_{\tilde{e}} + \sigma_{Z\tilde{e}}
\right) \frac{1}{2}(2 - \delta_{ij})
\end{equation}
is derived from the Lagrangians in App.~\ref{lagrangians}.

For beams with longitudinal polarization
$P_3^-$ for electrons and $P_3^+$ for positrons ($-1<P_3^\mp<1$)
one obtains for the $s$ channel contribution
\begin{eqnarray}
\sigma_Z & = & \frac{g_2^4}{32\pi \cos^4\theta_W} \frac{w_{ij}}{s^2}
  \nonumber \\
 && \times \Bigg\{ \hspace{2.5mm} \sum_{n=1}^{n_Z} \Big[ |D_{Z_n}(s)|^2
  (O^{''nL}_{ij})^2 \nonumber \\[-3mm]
 && \hspace{20mm} \times  \left[(1-P_3^- P_3^+)(L_n^2 + R_n^2)
  + (P_3^- - P_3^+) (R_n^2 - L_n^2)\right] \Big] \nonumber \\[3mm]
 && \hspace{5mm}{}+ \!\!\sum_{n,n'=1 \atop n<n'}^{n_Z}
  \Big[2 \, \mathrm{Re}[D_{Z_n}(s) D_{Z_{n'}}^*(s)] O^{''nL}_{ij}
  O^{''{n'}L}_{ij} \nonumber \\[-3mm]
 && \hspace{13mm} \times \left[(1 - P_3^- P_3^+) (L_n L_{n'} + R_n R_{n'})
  + (P_3^- - P_3^+) (R_n R_{n'} - L_n L_{n'})\right]\Big]\Bigg\}
     \nonumber \\[2mm]
 && \times \left\{ s^2 - (m_{\tilde{\chi}^0_i}^2 -
  m_{\tilde{\chi}^0_j}^2)^2 + \frac{1}{3} w_{ij}^2 - 4
  \eta_{\tilde{\chi}^0_i} \eta_{\tilde{\chi}^0_j}
  m_{\tilde{\chi}^0_i} m_{\tilde{\chi}^0_j} s \right\} . \label{twq1}
\end{eqnarray}
The $t$  and $u$ channel terms read
\begin{eqnarray}
\sigma_{\tilde{e}} & = & \frac{g_2^4}{32\pi} \frac{w_{ij}}{s^2}
 \nonumber \\[1mm]
 && \times \Bigg\{ (f^L_{ei})^2 (f^L_{ej})^2
  \left[(1 - P_3^- P_3^+) - (P_3^- - P_3^+) \right] \nonumber \\
 && \hspace{10mm} \times \bigg[ \frac{s^2 - (m_{\tilde{\chi}^0_i}^2 -
  m_{\tilde{\chi}^0_j}^2)^2 - 4 s^2 d_L (1 - d_L)}{4 s^2 d_L^2 - w_{ij}^2} + 1
  \nonumber\\
 && \hspace{15mm} {}+ \frac{2 s d_L - s +
  \eta_{\tilde{\chi}^0_i} \eta_{\tilde{\chi}^0_j}
  m_{\tilde{\chi}^0_i} m_{\tilde{\chi}^0_j}/d_L}{w_{ij}}
  \ln \bigg|\frac{2sd_L-w_{ij}}{2sd_L+w_{ij}}\bigg| \bigg] \Bigg\}
  \nonumber\\[5mm]
 && {}+ (L \rightarrow R,\quad P_3^- \rightarrow P_3^+)
     \label{twq2}
\end{eqnarray}
and the interference terms are
\begin{eqnarray}
\lefteqn{\sigma_{Z\tilde{e}} =}  \nonumber \\
 && \frac{g_2^4}{32\pi \cos^2\theta_W}
  \frac{w_{ij}}{s^2} \nonumber \\[1mm]
 && \times \Bigg\{ \sum_{n=1}^{n_Z} \left[\mathrm{Re}( D_{Z_n}(s) )
  O^{''nL}_{ij} L_n \right]
  f^L_{ei} f^L_{ej}
  \left[(1 - P_3^- P_3^+) - (P_3^- - P_3^+) \right] \nonumber \\
 && \hspace{7mm} \times \bigg[ \frac{s^2 - (m_{\tilde{\chi}^0_i}^2 -
  m_{\tilde{\chi}^0_j}^2)^2 - 4
  \eta_{\tilde{\chi}^0_i} \eta_{\tilde{\chi}^0_j}
  m_{\tilde{\chi}^0_i} m_{\tilde{\chi}^0_j} s - 4 s^2 d_L (1-d_L)}{w_{ij}} \ln
  \bigg|\frac{2sd_L-w_{ij}}{2sd_L+w_{ij}}\bigg| \nonumber \\
 && \hspace{13mm} {}- 4s (1 - d_L) \bigg] \Bigg\} \nonumber \\[2mm]
 && {}- ( L \rightarrow R,\quad P_3^- \rightarrow P_3^+)
  . \label{twq3}
\end{eqnarray}
The following abbreviations have been used
\begin{eqnarray}
D_{Z_n}(s) & \equiv & \frac{1}{s-m^2_{Z_n}+i m_{Z_n} \Gamma_{Z_n}} \,,\\
w_{ij} & \equiv & \left[s - (m_{\tilde{\chi}^0_i}+m_{\tilde{\chi}^0_j})^2\right]
^{\frac{1}{2}}
  \left[s - (m_{\tilde{\chi}^0_i}-m_{\tilde{\chi}^0_j})^2\right]^{\frac{1}{2}} ,
  \\
d_{L,R} & \equiv & \frac{1}{2s}\left( s + 2 m_{\tilde{e}_{L,R}}^2 -
  m_{\tilde{\chi}^0_i}^2 - m_{\tilde{\chi}^0_j}^2 \right) .
\end{eqnarray}

For $n_Z=1$ one recovers the cross section of the MSSM
\cite{bartlfraasneutprod}, if the couplings $O^{''1L}_{ij} \equiv
O^{''L}_{ij}$, $L_1 \equiv L$, $R_1 \equiv R$ and $f^{L/R}_{ei}$ are
interpreted as those of the MSSM. 
In models R5$_1$ and R5$_2$ the number of neutral gauge bosons is
$n_Z=2$, in model R6 $n_Z=3$.
Note that all
couplings are assumed to be real due to $CP$ conservation.

Finally we define the polarization asymmetry
\begin{equation} \label{AeDef}
  A_\mathrm{LR} = \frac{\sigma(-P_3^-,P_3^+ ) - \sigma(+P_3^-,P_3^+ )}{
    \sigma(-P_3^-,P_3^+) + \sigma(+P_3^-,P_3^+)} \, .
\end{equation}
with respect to the electron polarization $P_3^-$ and fixed positron
polarization $P_3^+$. In the following numerical discussions of
$A_\mathrm{LR}$ we use $P_3^- = 0.85$ \cite{gudipol}.

\subsection{Scenarios}

Neutralino production will be discussed in representative mixing
scenarios in the extended models R5$_1$, R5$_2$, and R6 (see
Tables~\ref{parameter} -- \ref{szenr52}).

The experimental lower mass bounds on the new \es\ gauge bosons of
about 600~GeV \cite{abe1} are respected by choosing a
value of 3000~GeV for the vacuum expectation values
of the singlet fields $v_3$ and $v_4$ which leads to  
$m_{Z_2} = 1264$~GeV in the model R5$_1$, $m_{Z_2} =
1786$~GeV in R5$_2$ and $m_{Z_2} = 1383$~GeV, $m_{Z_3} = 1786$~GeV
in R6.
The widths of the new gauge bosons are estimated by
$\Gamma_{Z_{2,3}} \approx 0.014 \: m_{Z_{2,3}}$
\cite{gherghetta}.


The neutralino mixing parameters $M_2$,
$\lambda$, $\tan\beta=v_2/v_1$, $v_3$ $[\textrm{and }v_4]$
are fixed in Table~\ref{parameter} for three scenarios H, G, M
where the light MSSM-like neutralinos have higgsino, gaugino
and mix character, respectively,
and the mass of the lightest MSSM-like neutralino is 100 GeV.
The higgsino mass parameter $\mu$ of the MSSM is recovered by
$\mu=-\lambda v_3$.
Since the neutralino production cross sections depend only
weakly on $\tan\beta$,
we confine ourselves to one value $\tan\beta=5$.

The neutralino masses and mixings of Table~\ref{szenmssm} are obtained
assuming the GUT relation \cite{gunion, bartlfraasneut}
\begin{equation}
  M_1 = M' \:[=M''] = M_2 \frac{5}{3}\tan^2\theta_W \,.
\end{equation}
Then the light neutralinos are MSSM-like in models
R5$_1$ and R6.

In Table~\ref{szenr51}, however, we abandon the GUT-relation for $M'$
and choose large values which lead to light neutralinos with singlino character
\cite{decarlosespinosa, suematsu3}.
Here a singlino-like $\tilde{\chi}^0_1$ with mass of about 80~GeV
appears in models R5$_1$ and R6.
The $\tilde{\chi}^0_1$ in the model R5$_2$ is always very light with mass
$\mathcal{O}(0.1~\mathrm{GeV})$ as shown in Table~\ref{szenr52}.

The neutralino cross sections in $e^+e^-$ annihilation also depend on
the masses of the left and right selectrons.
In order to compare the results, we use the same
weak scale selectron masses throughout our numerical analysis.
First the mass of the left selectron is fixed at $m_{\tilde{e}_L} = 300$ GeV in
both the MSSM and the \es\ models. Then one obtains a right selectron
mass $m_{\tilde{e}_R} = 200$~GeV by
mSUGRA-type renormalization group equations with parameters $M_2 =
300$~GeV and $m_0 = 132$~GeV in the MSSM \cite{mselMSSM}.
Assuming $\tilde{M}_{\tilde{e}_L}^2 = \tilde{M}_{\tilde{e}_R}^2$ in
the selectron mass formulas in the \es\ models (App.~\ref{sfermionmasses})  
the large D-terms of the new U(1) gauge factors result in
$m_{\tilde{e}_R} = 753$~GeV in R5$_1$
and $m_{\tilde{e}_R} = 1022$~GeV in R5$_2$ and R6.
Note that in the \es\ models in contrast to the MSSM the
right selectrons are much heavier than the left ones \cite{drees,MsfRGEsammel}.


In the scenarios without GUT relations we keep the above values for
the selectron masses as free parameters.
Otherwise the RGE would induce weak scale
selectron masses of the order of $M'$ which strongly suppress the
cross sections especially in gaugino scenarios.

\begin{table}[ht]
\renewcommand{\arraystretch}{1.8}
\centering
\begin{tabular}{||l||c|c|c||} \hline\hline
  Scenario  &  \makebox[15mm]{H} & \makebox[15mm]{G} &
  \makebox[15mm]{M} \\ \hline \hline 
  $M_2/$GeV & $400$ & $-209$ & $-251$ \\ \hline
  $\lambda$ & $0.037$ & $0.133$ & $0.058$ \\ \hline
  $M_1$ & \multicolumn{3}{c||}{$\disp M_2 \: 5/3 \tan^2 \theta_W$} \\ \hline
  $\tan\beta$ & \multicolumn{3}{c||}{$5$} \\ \hline
  $v_3\:[= v_4]/$GeV & \multicolumn{3}{c||}{$3000$} \\ \hline\hline
\end{tabular}
\caption{\label{parameter} Parameters of the neutralino mixing scenarios
  in the \es\ models and in the MSSM with $\mu = - \lambda v_3$.} 
\end{table}

\begin{table}[p]
\renewcommand{\arraystretch}{1.8}
\centering
\begin{tabular}{||l||c|c|c||} \hline\hline
  Model & \multicolumn{3}{c||}{MSSM, R5$_1$, R6
  ($M'\:[=M''] = M_1$)} \\ \hline \hline
  Scenario  &  \makebox[20mm]{H} & \makebox[20mm]{G} &
  \makebox[20mm]{M} \\ \hline \hline 
  $m_{\tilde{\chi}^0_1}/$GeV & $100$ & $100$ &  $100$ \\ \hline
  $\tilde{\chi}^0_1$-character & higgsino & gaugino & mix \\ \hline
  $m_{\tilde{\chi}^0_2}/$GeV & $124$ & $192$ & $161$ \\ \hline
  $\tilde{\chi}^0_2$-character & higgsino & gaugino & mix \\ \hline \hline
\end{tabular}
\caption{\label{szenmssm} Neutralino masses and mixings in the MSSM
with $\mu = - \lambda v_3$ and in the \es\ models R5$_1$ and R6 
with $M'\:[=M''] = M_1$.}
\end{table}

\begin{table}[p]
\renewcommand{\arraystretch}{1.8}
\centering
\begin{tabular}{||l||c|c|c||} \hline\hline
  Model & \multicolumn{3}{c||}{R5$_1$ ($M'=20$ TeV), R6 ($M'=32$ TeV)}
  \\ \hline \hline
  Scenario  &  \makebox[22mm]{H} & \makebox[22mm]{G} &
  \makebox[22mm]{M} \\ \hline \hline 
  $m_{\tilde{\chi}^0_1}/$GeV & $80$ & $80$ &  $80$ \\ \hline
  $\tilde{\chi}^0_1$-character & singlino & singlino & singlino \\ \hline
  $m_{\tilde{\chi}^0_2}/$GeV & $100$ & $100$ & $100$ \\ \hline
  $\tilde{\chi}^0_2$-character & higgsino & gaugino & mix \\ \hline \hline
\end{tabular}
\caption{\label{szenr51} Neutralino masses and mixings in the
\es\ models R5$_1$ with $M'=20$ TeV and R6 with $M'=32$ TeV, $M''=M_1$.}
\end{table}

\begin{table}[p]
\renewcommand{\arraystretch}{1.8}
\centering
\begin{tabular}{||l||c|c|c||} \hline\hline
  Model & \multicolumn{3}{c||}{R5$_2$ ($M'=M_1$)} \\ \hline \hline
  Scenario  &  \makebox[16mm]{H} & \makebox[16mm]{G} & \makebox[16mm]{M} \\ \hline \hline
  $m_{\tilde{\chi}^0_1}/$GeV & $0.1$ & $0.2$ & $0.1$ \\ \hline
  $\tilde{\chi}^0_1$-character & singlino  & singlino & singlino \\ \hline
  $m_{\tilde{\chi}^0_2}/$GeV & $100$ & $100$ & $100$ \\ \hline
  $\tilde{\chi}^0_2$-character & higgsino & gaugino & mix \\ \hline \hline
\end{tabular}
\caption{\label{szenr52} Neutralino masses and mixings
  in the \es\ model R5$_2$ with $M'=M_1$.}
\end{table}

\subsection{Numerical results}

\subsubsection{\boldmath$\tilde{\chi}^0_1$ and $\tilde{\chi}^0_2$ MSSM-like}

\begin{figure}[p]
\centering
\begin{picture}(16,18.3)
  \put(0,-.05){\epsfig{file=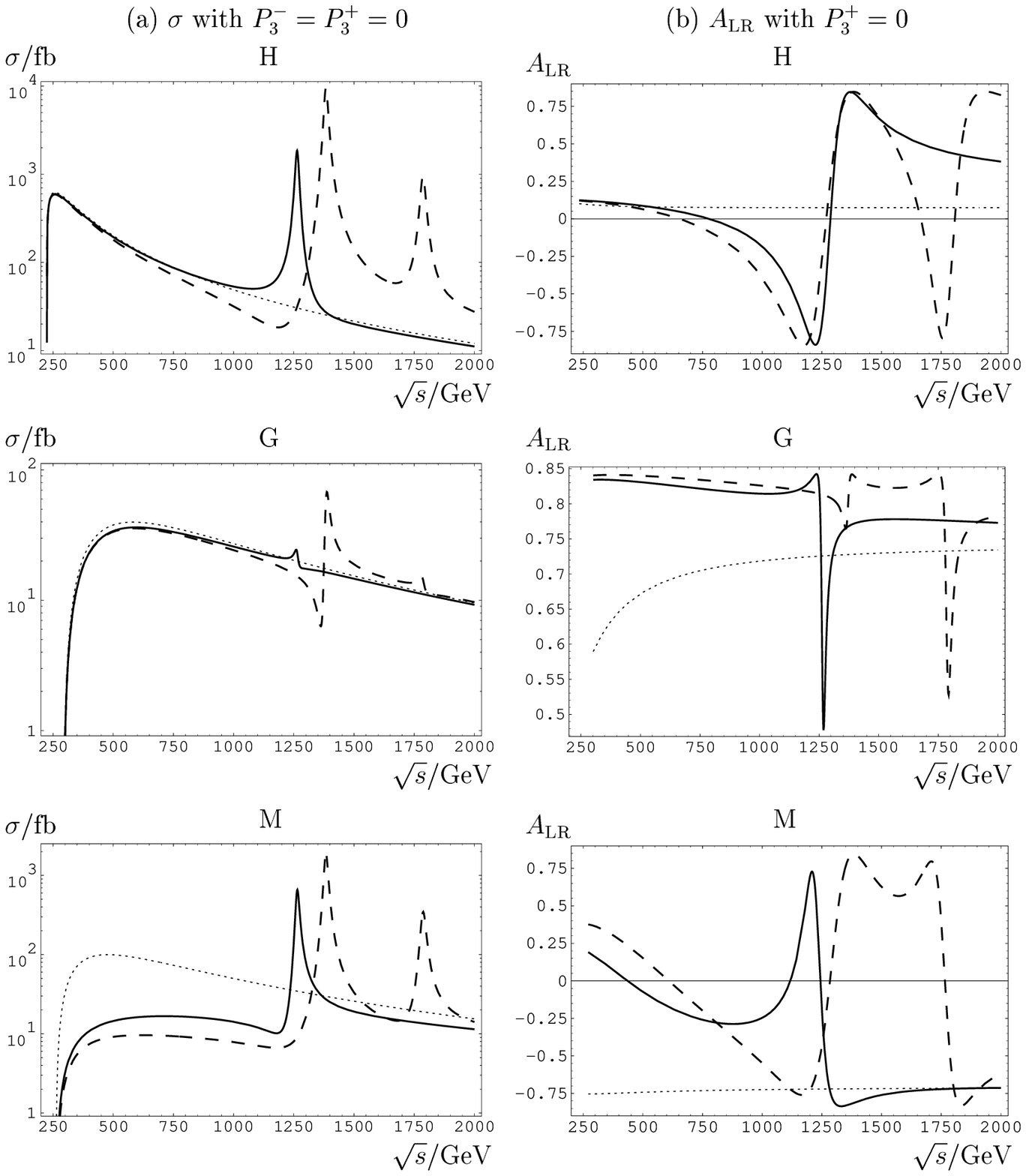}}
\end{picture}
\caption{\label{totwq12mssm} (a) Total cross sections for $P_3^- =
  P_3^+ = 0$ and  (b) polarization asymmetries for $P_3^+ = 0$
  of the process $e^+e^- \to \tilde{\chi}^0_1 \tilde{\chi}^0_2$ in the
  scenarios of Table~\ref{parameter} with $M'\:[=M''] = M_1$ 
  in the models R5$_1$ (solid), R6 (dashed) and the MSSM
  (dotted).}
\end{figure}

In the models R5$_1$ and R6
with $M' \; [= M''] = M_1$ both light neutralinos
$\tilde{\chi}^0_1$ and $\tilde{\chi}^0_2$ have MSSM-like
character in all scenarios 
of Table~\ref{parameter}
with masses given in Table~\ref{szenmssm}.
The total cross sections for neutralino production
$e^+e^- \to \tilde{\chi}^0_1 \tilde{\chi}^0_2$
with unpolarized
beams ($P_3^- = P_3^+ = 0$) and the polarization asymmetries for
$P_3^+ = 0$ are shown
in Fig.~\ref{totwq12mssm}. 

In scenario H with higgsino-like neutralinos
the new gauge bosons cause high narrow resonances, 
whereas otherwise the cross sections are 
similar to the MSSM. The polarization asymmetries, however, show a
much wider resonance effect. Contrary to the MSSM, where
the electron polarization asymmetry
$A_\mathrm{LR} \sim 0.1$ is nearly independent of the beam energy,
it changes sign at about 800~GeV in model R5$_1$ 
and at about 650~GeV in model R6. 

We do not explicitly show polarization asymmetries for $P_3^+ \neq 0$
in scenario H since
additional polarization of the positron beam only shifts the asymmetry
in all models for $P_3^+>0$ to higher or for $P_3^+<0$ to lower values.
In the \es\ models one obtains at threshold 
$A_\mathrm{LR} \sim 0.6$
for $P_3^+ = +0.6$ 
and $A_\mathrm{LR} \sim -0.4$ for $P_3^+ = -0.6$, the change
of sign of $A_\mathrm{LR}$ occurs for $P_3^+ = +0.6$ at
$\sqrt{s}=1150$~GeV (R5$_1$) or 
$\sqrt{s}=1050$~GeV (R6) and for $P_3^+ = -0.6$ at the $Z'$ resonance
significantly above the energy range
of a linear collider at first stage.

In scenario G the gaugino-like light neutralinos are mainly produced 
by the exchange of left selectrons 
leading to obviously smaller gauge boson resonances.
Choosing the same left selectron mass in the MSSM and the \es\ models 
the cross sections are rather similar.
The different masses of the right
selectrons, however, lead to distinct differences between
the electron polarization asymmetries that are largest just above
threshold \cite{christova}
where $A_\mathrm{LR} \sim 0.84$ in the \es\ models and $A_\mathrm{LR} \sim
0.59$ in the MSSM.

\begin{figure}[ht]
\centering
\begin{picture}(16,5.8)
  \put(0,-.05){\epsfig{file=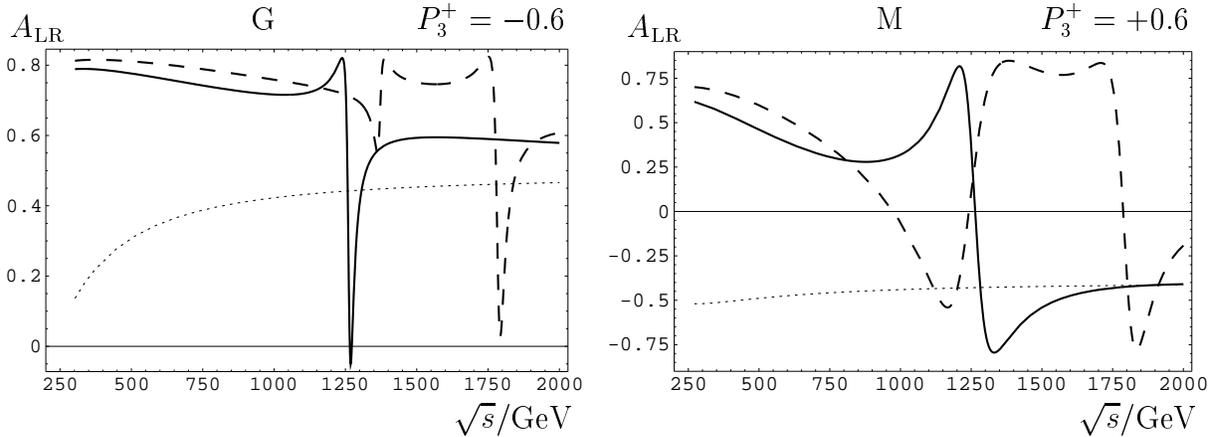}}
\end{picture}
\caption{\label{asyme12mssmPOL} Polarization asymmetries for polarized
  positron beam of the process $e^+e^- \to \tilde{\chi}^0_1
  \tilde{\chi}^0_2$ in the scenarios G and M of Table~\ref{parameter}
  with $M'\:[=M''] = M_1$
  in the models R5$_1$ (solid), R6 (dashed) and the MSSM
  (dotted).}
\end{figure}

Positron polarization hardly affects the polarization asymmetries in
the \es\ models whereas the asymmetry in the MSSM is shifted to 
lower (higher) values for $P_3^+<0\;(>0)$.
Fig.~\ref{asyme12mssmPOL} shows how 
negative positron beam polarization
$P_3^+ = -0.6$ 
enhances the differences between the models
($A_\mathrm{LR} \sim 0.8$ at threshold in the
\es\ models compared to $A_\mathrm{LR} \sim 0.15$ in the MSSM).

In scenario M the gaugino content of the
light neutralinos $\tilde{\chi}^0_1$ and $\tilde{\chi}^0_2$ 
couples mainly to right selectrons which are much heavier in the
\es\ models than in the MSSM. The different right selectron
masses lead to significantly smaller cross sections in the \es\ models
below the resonances compared to the MSSM
and clearly distinguishable electron polarization asymmetries at threshold
($A_\mathrm{LR} \sim
0.4$ in the model R6, $0.2$ in the model R5$_1$
and $-0.75$ in the MSSM).
This effect can be used to determine the selectron masses \cite{ustronpaper}.

Due to the higgsino content of the neutralinos,
the cross sections
show high narrow resonances as in scenario H 
with wider resonances in $A_\mathrm{LR}$ than in scenario G.
Positive positron beam polarization shifts the asymmetries to
higher values in all models but in a larger extend in the
\es\ models than in the MSSM
(see Fig.~\ref{asyme12mssmPOL}).
The threshold values $A_\mathrm{LR} \sim 0.7$ in model R6,
$0.6$ in model R5$_1$ and $-0.5$ in the MSSM for $P_3^+ = +0.6$ 
offer a clear signature to distinguish between the models.

\subsubsection{\boldmath$\tilde{\chi}^0_1$ singlino-like and $\tilde{\chi}^0_2$
  MSSM-like}

\begin{figure}[ht!]
\centering
\begin{picture}(16,5.8)
  \put(0,-.05){\epsfig{file=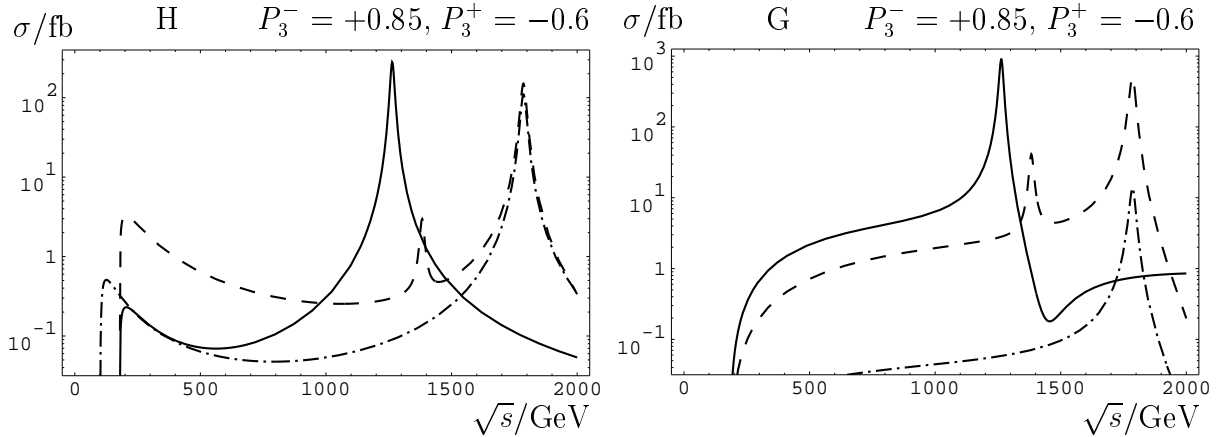}}
\end{picture}
\caption{\label{totwq12singlet} Total cross sections for $e^+e^- \to
  \tilde{\chi}^0_1 \tilde{\chi}^0_2$ in the scenarios H and G 
  of Table~\ref{parameter} with beam polarization $P_3^- = +
  0.85$, $P_3^+ = - 0.6$ 
  in the models R5$_1$ with $M'=20$~TeV (solid), R5$_2$
  with $M'=M_1$ (dashed-dotted) and R6
  with $M'=32$~TeV (dashed).}
\end{figure}

In the models R5$_1$ with $M'=20$~TeV, R5$_2$
with $M'=M_1$ and R6 with $M'=32$~TeV 
the lightest neutralino $\tilde{\chi}^0_1$ is a nearly
pure singlino. Then the second lightest neutralino has similar 
higgsino, gaugino or mix character in the scenarios H, G or M, 
respectively, as the $\tilde{\chi}^0_1$ in the MSSM
(Tables~\ref{szenr51} and \ref{szenr52}).
Since the singlino content of the neutralinos does not couple to
selectrons and the standard gauge boson 
the direct production of singlino dominated neutralinos
is generally suppressed.
Nevertheless Fig.~\ref{totwq12singlet} shows 
total cross sections for a beam polarization $P_3^- = + 0.85$ and
$P_3^+ = - 0.6$ 
that 
reach some 0.1~fb outside the gauge boson resonances
in all models in scenario H and in models R5$_1$ and R6 in scenario G.
So even the direct production of singlino-like neutralinos
may be detectable in a considerable domain of the parameter space
at a linear collider with a luminosity of
500~$\mathrm{fb}^{-1}$ at $\sqrt{s}=500$~GeV \cite{sitges,LCTH2000}.

The size of the cross sections in the different models mainly depends
on the MSSM-components of the singlino dominated 
$\tilde{\chi}^0_1$. In scenario H the doublet 
higgsino content ($N_{13}^2 + N_{14}^2$), which
couples to the doublet higgsino-like $\tilde{\chi}^0_2$, is $0.22\;\%$
in the model R5$_1$, $0.17\;\%$ in R5$_2$
and $0.42\;\%$ in R6.
For $\sqrt{s}\lesssim 1$~TeV the polarized cross sections with
$P_3^- = + 0.85$ and $P_3^+ = - 0.6$ are about $1.5$ to 3 times 
larger than the unpolarized cross sections depending on the model.

In scenario G the singlino dominated $\tilde{\chi}^0_1$ has
a MSSM-gaugino content ($N_{11}^2 + N_{12}^2$)
of $1.5\;\%$
in R5$_1$, $0.05\;\%$ in R5$_2$
and $1.7\;\%$ in R6. Therefore the
cross section in R5$_2$ is smaller than
0.1~fb outside the resonance while the larger cross section 
in R5$_1$ compared to R6 is caused by the smaller 
mass of the right selectron.
The beam polarization enhances the unpolarized cross sections 
by a factor of about 3.

\begin{figure}[ht!]
\centering
\begin{picture}(16,5.8)
  \put(0,-.05){\epsfig{file=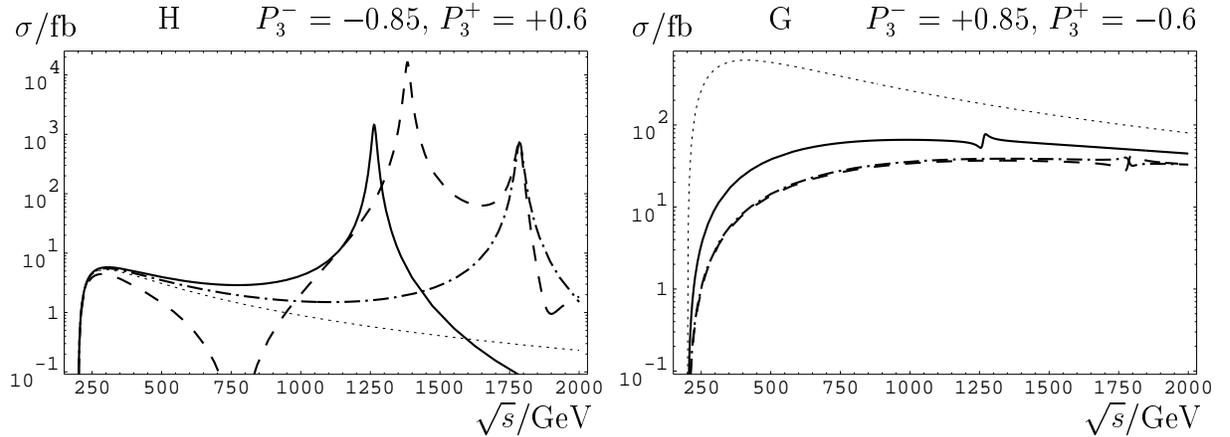}}
\end{picture}
\caption{\label{totwq22singlet} Total cross sections for $e^+e^- \to
  \tilde{\chi}^0_2 \tilde{\chi}^0_2$ in scenario H  
  with beam polarization $P_3^- = -
  0.85$, $P_3^+ = + 0.6$ and scenario G with 
  $P_3^- = + 0.85$ and $P_3^+ = - 0.6$ 
  in the models R5$_1$ with $M'=20$~TeV (solid), R5$_2$
  with $M'=M_1$ (dashed-dotted) and R6
  with $M'=32$~TeV (dashed).
  For comparison the corresponding invisible process
  $e^+e^- \to \tilde{\chi}^0_1 \tilde{\chi}^0_1$ in the MSSM is
  plotted (dotted).}
\end{figure}

The cross sections for the pair
production of the lightest MSSM-like neutralinos
$e^+ e^- \to \tilde{\chi}^0_2 \tilde{\chi}^0_2$ 
are plotted in Fig.~\ref{totwq22singlet}. 
The corresponding process $e^+e^- \to \tilde{\chi}^0_1 \tilde{\chi}^0_1$
in the MSSM is invisible.

In scenario H a beam polarization $P_3^- = -0.85$, $P_3^+ = + 0.6$
roughly doubles the unpolarized 
cross sections to about 4 to 6~fb for $\sqrt{s} \lesssim
1$~TeV. Pair production of higgsino-like neutralinos is generally
suppressed compared to the associated production of higgsino
dominated neutralinos in Fig.~\ref{totwq12mssm}.
The minimum of the cross section in model R6 at
$\sqrt{s} \approx 750$~GeV is caused by negative interference effects
between the contributions of the three gauge bosons.

In scenario G the opposite beam polarization $P_3^- = +0.85$ and $P_3^+
= - 0.6$ leads to a maximum enhancement of the
unpolarized cross sections by a factor between 2 and 3.
Since the gaugino-like $\tilde{\chi}^0_2$ couples mainly to
right selectrons in the whole parameter space
also the cross section for the pair
production of gaugino-like neutralinos is suppressed in the \es\ models 
compared to the MSSM.

The polarized cross sections in scenarios H and G are clearly above the
discovery limit of a high luminosity linear collider.
An even more distinctive signal can be expected
from the process $e^+e^- \to \tilde{\chi}^0_2  
\tilde{\chi}^0_3$ with a cross section similar to the cross section of $e^+e^- \to
\tilde{\chi}^0_1 \tilde{\chi}^0_2$ for MSSM-like
$\tilde{\chi}^0_1$ and $\tilde{\chi}^0_2$ (see Fig.~\ref{totwq12mssm}). 

\section{Conclusion}

We have analyzed the neutralino sector in \es\ inspired extended
supersymmetric models with additional neutral gauge
bosons and singlet Higgs fields.
To obey the experimental lower mass bounds of the new gauge bosons
the vacuum expectation values of the singlet fields must be of the
order of some TeV. 

In a rank-5 model with two singlets the lightest
neutralino is always a very light nearly pure singlino.
Light neutralinos with singlino character also appear in a
rank-5 model with one singlet and in the rank-6 model 
if the $\mathrm{U(1)'}$ ($\mathrm{U(1)''}$)
gaugino mass parameter $M'$ ($M''$) takes large values
$\mathcal{O}(10~\mathrm{TeV})$ because of a see-saw-like mechanism in
the submatrix of the exotic neutralinos. 
Two light singlino-like neutralinos may exist 
in the rank-5 model with
two singlets for large $M'$ 
and in the rank-6 model for both $M'$ and $M''$ large.
However, light neutralinos in the discussed \es\ models
never have dominant $Z'$ ($Z''$) gaugino character.
Assuming the GUT relation for the gaugino mass parameters
in the rank-5 model with one singlet and the rank-6 model, the
MSSM-like neutralinos decouple and the masses of the exotic neutralinos
are of the order of the singlet vacuum expectation values in the TeV range.

The production of neutralinos in $e^+e^-$ annihilation proceeds via
$s$ channel exchange of neutral gauge bosons and $t$ and $u$ channel
exchange of selectrons. 
The production cross sections of neutralinos with dominant higgsino
character show narrow high
resonances of the new gauge bosons but are otherwise rather similar to
the MSSM.
If the resonances of the new gauge bosons are not accessible
at the first stage of a linear collider,
the use of polarized beams is an important tool to discriminate 
between the MSSM and the \es\ models  
since the polarization asymmetries show
significantly wider resonance effects
far below the mass of the new gauge bosons.

Assuming mSUGRA-type RGEs the models also differ by the selectron masses.
Due to additional D-terms in the superpotential
the right selectrons
are much heavier than the left ones in the \es\ models contrary to the MSSM.
Then in scenarios with large gaugino content of the neutralinos the
differences between the models depend on the neutralino couplings to 
left and right selectrons.
Especially if one or both selectrons cannot be produced directly at the first
stage of a linear collider
the determination of the selectron masses by measuring neutralino
production cross sections 
offers a particularly suitable possibility to distinguish between the models.
Polarization asymmetries show even more distinctive 
effects of the selectron masses.

The cross sections for the direct production of light singlino-like
neutralinos are typically of the order of some fb outside the 
gauge boson resonances
which is sufficient to be detected at a high
luminosity linear collider. If the lightest supersymmetric particle
is a singlino-like neutralino, 
pair production of $\tilde{\chi}^0_2$ with a significant gaugino content 
is visible while pair production of a higgsino-dominated neutralino is 
generally suppressed.
Again, beam polarization enhances the cross sections by a factor up to 3
and improves the discovery chances at a linear collider.

\section*{Acknowledgment}

We are indebted to G.~Moortgat-Pick for the excellent collaboration over
many years.
S.H. is grateful to the Department of Physics of the University
of Wisconsin at Madison for the kind hospitality and the pleasant atmosphere.
S.H. is supported by the Deutsche Forschungsgemeinschaft (DFG) under
contract No.\ HE~3241/1-1.
F.F. and H.F. are supported
by the Deutsche Forschungsgemeinschaft (DFG) under contract No.\
\mbox{FR~1064/4-1},
by the Bundesministerium f\"ur
Bildung und Forschung (BMBF) under contract No.
05~HT9WWA~9 and by
the Fonds zur F\"orderung der wissenschaftlichen For\-schung of Austria,
project No.~P13139-PHY.

\section*{Appendix: E$_\mathbf{6}$ inspired models}

\renewcommand{\theequation}{\thesection.\arabic{equation}}

\begin{appendix}

\section{Symmetry breaking \label{e6breaking}}
\setcounter{equation}{0}

The exceptional group \es\ may be a suitable candidate for a
gauge group of a Grand Unified Theory (GUT). The
rank-6 group \es\ is a natural extension of the rank-4 group SU(5) and
the rank-5 group SO(10) and contains the maximal subgroup
$\mathrm{SO(10) \times U(1)}$. \es\ holds complex representations
necessary to describe chiral fermions and is naturally free of
anomalies. Furthermore the compactification of an $\mathrm{E_8 \times
E_8'}$ string theory to four dimensions can lead to \es\ as an
effective GUT group \cite{hr, boyce, e6math, london}.

In order to get a low energy gauge group of the form $\mathrm{SU(3)
\times SU(2) \times U(1)}^n$ the \es\ group has to be
broken \cite{hr}. If \es\
is broken directly to a low energy group of rank 5 these group
\begin{equation} \label{egrr5}
  \mathrm{SU(3)_C \times SU(2)_L \times U(1)_\mathit{Y} \times
  U(1)_\eta}
\end{equation}
is uniquely determined. If \es\ is broken to a low energy group of
rank 6 several possibilities arise. We confine
ourselves on the case of two additional U(1) factors in comparison to
the SM
\begin{equation} \label{egrr6}
  \mathrm{SU(3)_C \times SU(2)_L \times U(1)_\mathit{Y} \times
  U(1)_\psi \times U(1)_\chi} \,,
\end{equation}
where $\mathrm{U(1)_\psi}$ and $\mathrm{U(1)_\chi}$ are defined by
\begin{equation}
  \mathrm{E_6} \rightarrow \mathrm{SO(10) \times U(1)_\psi}\,,
  \qquad
  \mathrm{SO(10)} \rightarrow \mathrm{SU(5) \times U(1)_\chi}\,.
\end{equation}

For suitable large vacuum expectation values of the symmetry breaking
Higgs fields the rank-6 model can be reduced to an ``effective''
rank-5 model ($\mathrm{U(1)_\psi \times U(1)_\chi \to U(1)_\theta}$),
where one new gauge boson decouples from low energy theory
\cite{hr, london}.
Then the remaining new gauge boson $Z' = Z_\psi \cos\theta - Z_\chi
\sin\theta$ is in general a mixture of $Z_\psi$ and $Z_\chi$. For
$\theta = \arcsin \sqrt{3/8}$ the quantum numbers of the true rank-5
model (``model $\eta$'') are recovered. In this paper we
focus on this model $\eta$ in the rank-5 case.
$\theta = -\arctan\sqrt{1/15}$ gives the so called U(1)$_N$ model which is
favored by neutrino phenomenology and leptogenesis considerations
\cite{raidal}.

\section{Particle content\label{partcont}}
\setcounter{equation}{0}

Each chiral generation of fermions belongs to a
fundamental representation $\mathbf{27}$
of \es\ which decomposes according to
\begin{equation} \label{grpzerl}
  \mathbf{27} = (\mathbf{16},\mathbf{10}) +
  (\mathbf{16},\mathbf{\bar{5}}) +  (\mathbf{16},\mathbf{1}) +
  (\mathbf{10},\mathbf{5}) + (\mathbf{10},\mathbf{\bar{5}}) +
  (\mathbf{1},\mathbf{1})
\end{equation}
in terms of the subgroups SO(10) and SU(5) of \es\ \cite{hr}.
Table \ref{quantnr} shows this for one generation of matter in the \es\
models.
In order to fill the $\mathbf{27}$ new ``exotic'' fields are necessary in
comparison to the MSSM. The breaking of \es\ fixes the color, isospin and
hypercharge of these ``exotics'' but not their baryon and lepton numbers
and their $R$-parity.
In this paper we assume vanishing baryon and lepton numbers and $R$-parity
$-1$ for the fermions $H$, $H^c$, $\nu^c_L$ and $S^c_L$. So they can be
interpreted as superpartners of doublet and singlet Higgs fields of the model.

\begin{table}[ht!]
\renewcommand{\arraystretch}{1.5}
\begin{center}
\begin{tabular}{||ll@{\hspace{1mm}}c@{\hspace{1mm}}ccccc||}
 \hline\hline
 &  & $\mathrm{SU(3)_C}$ & $T_{3L}$ & $Y$ & $Y_\eta$ & $Y_\psi$ &
  $Y_\chi$ \\ \hline\hline
\rule{0mm}{8mm}$\disp Q\equiv{u \choose d}_L$ &
  $(\mathbf{16},\mathbf{10})$ & $\mathbf{3}$ &
  $\disp {+1/2 \choose -1/2}$ & $1/3$ & $2/3$ & $\sqrt{10}/6$ &
  $-1/\sqrt{6}$  \\
$u^c_L$ & & $\mathbf{\bar{3}}$ & 0 & $-4/3$ & $2/3$ & $\sqrt{10}/6$ &
  $-1/\sqrt{6}$ \\
$e^c_L$ & & $\mathbf{1}$ & 0 & $2$ & $2/3$ & $\sqrt{10}/6$ &
  $-1/\sqrt{6}$ \\ \hline
\rule{0mm}{8mm}$\disp L\equiv{\nu \choose e}_L$ &
  $(\mathbf{16},\mathbf{\bar{5}})$
  & $\mathbf{1}$ &
  $\disp {+1/2 \choose -1/2}$ & $-1$ & $-1/3$ & $\sqrt{10}/6$ &
  $3/\sqrt{6}$ \\
$d^c_L$ & & $\mathbf{\bar{3}}$ & 0 & $2/3$ & $-1/3$ & $\sqrt{10}/6$ &
  $3/\sqrt{6}$ \\ \hline
$\nu^c_L$ & $(\mathbf{16},\mathbf{1})$ & $\mathbf{1}$ & 0 & 0 & $5/3$
  & $\sqrt{10}/6$ &
  $-5/\sqrt{6}$ \\ \hline\hline
\rule{0mm}{8mm}$\disp H\equiv{N \choose E}_L$ &
  $(\mathbf{10},\mathbf{\bar{5}})$ & $\mathbf{1}$ &
  $\disp {+1/2 \choose -1/2}$ & $-1$ & $-1/3$ & $-\sqrt{10}/3$ &
  $-2/\sqrt{6}$ \\
$h^c_L$ & & $\mathbf{\bar{3}}$ & 0 & $2/3$ & $-1/3$ & $-\sqrt{10}/3$ &
  $-2/\sqrt{6}$ \\ \hline
\rule{0mm}{8mm}$\disp H^c \equiv{E \choose N}^c_L$ &
  $(\mathbf{10},\mathbf{5})$ &
  $\mathbf{1}$ &
  $\disp {+1/2 \choose -1/2}$ & $1$ & $-4/3$ & $-\sqrt{10}/3$ &
  $2/\sqrt{6}$ \\
$h_L$ & & $\mathbf{3}$ & 0 & $-2/3$ & $-4/3$ & $-\sqrt{10}/3$ &
  $2/\sqrt{6}$ \\ \hline\hline
$S^c_L$ & $(\mathbf{1},\mathbf{1})$ & $\mathbf{1}$ & 0 & 0 & $5/3$ &
  $2\sqrt{10}/3$ & 0 \\ \hline\hline
\end{tabular}
\caption{\label{quantnr}Fermionic particle content of the fundamental
$\mathbf{27}$ representation of \es, assignment of the fermions to the
subgroups SO(10), SU(5) and $\mathrm{SU(3)_C}$ and quantum numbers
according to $\mathrm{SU(2)_L}$, $\mathrm{U(1)}_Y$,
$\mathrm{U(1)}_\eta$, $\mathrm{U(1)}_\psi$, $\mathrm{U(1)}_\chi$
\cite{hr}.}
\end{center}
\end{table}

It is always possible to choose a basis that only the Higgs fields of one
generation get vacuum expectation values, conventionally the fields of the
third generation.
So the two doublet Higgs fields of the model $H_1$ and $H_2$ can be identified
as
\begin{equation} \label{dhiggs}
  H_1 \equiv (\tilde{H})_3 \textrm{\quad and\quad} H_2 \equiv
  (\tilde{H}^c)_3
\end{equation}
and two singlet Higgs fields as
\begin{equation} \label{shiggs}
  N_1 \equiv (\tilde{S}^c_L)_3 \textrm{\quad and\quad} N_2 \equiv
  (\tilde{\nu}^c_L)_3
\end{equation}
with the vacuum expectation values (vevs)
\begin{equation} \label{higgsvev}
  \langle H_1 \rangle = {v_1 \choose 0},\quad \langle H_2 \rangle =
    {0 \choose v_2},\quad \langle N_1 \rangle = v_3,\quad \langle N_2
    \rangle = v_4 \; .
\end{equation}
The corresponding fields of the first two generations which obtain no
vevs are called ``unHiggs'' \cite{hr} and are
discussed in detail in \cite{unhiggse}.
In particular the corresponding ``unhiggsinos'' do not mix with the
ordinary neutralinos.
However the mass of the lightest neutral unhiggsino has a strict upper
bound of about 100~GeV, hence it may be the LSP in some areas of the
parameter space \cite{dreestata}. 
This case has to be considered in the analysis of the neutralino
decay signatures,
but the results regarding the mass spectra
and the production of neutralinos discussed in this paper remain valid.

In the case of the rank-5 models only one singlet vev is necessary to
break the extended gauge group.
If the second singlet obtains no vev it decouples from the neutralino
sector and is also considered an unhiggsino (rank-5 model with one
singlet, R5$_1$).
This model also avoids the problems with the creation of a vev for the
second singlet \cite{hr,gunion,ellis,campbell}.

\section{Gauge boson sector\label{gaugesector}}
\setcounter{equation}{0}

Models with one (rank-5 models) or two (rank-6 models) U(1) factors in
the gauge group contain one ($Z'$) or two ($Z'$ and $Z''$) new neutral
gauge bosons in comparison to the SM. These new gauge bosons mix with
the standard $Z$ boson $Z^\mathrm{SM}$ to form mass eigenstates
$Z_1$, $Z_2$ [and $Z_3$].

In the rank-5 [rank-6] model with the electroweak gauge group
\begin{equation}
  \mathrm{SU(2)_L \times U(1)_\mathit{Y} \times U(1)'
     \,[\!{}\times U(1)'']}
\end{equation}
the mass term of the gauge bosons without abelian gauge kinetic
mixing reads
\begin{equation}
\mathcal{L}_M^\mathrm{Gauge} = \frac{1}{2} g_2^2 (v_1^2+v_2^2) W_\mu^+ W^{-\mu}
+ \frac{1}{2} \left( A_\mu , Z^\mathrm{SM}_\mu, Z'_\mu\,[, Z''_\mu] \right)
X \left( \begin{array}{c} A^\mu \\ {Z^\mathrm{SM}}^{\mu} \\ {Z'}^\mu \\
 {[{Z''}^\mu]}
\end{array}\right).
\end{equation}
As in the SM $W^\pm$ denote the charged gauge bosons with mass
\begin{equation} \label{mw}
  m_W^2=\frac{1}{2} g_2^2 (v_1^2+v_2^2).
\end{equation}
$X$ is the mixing matrix of the neutral gauge bosons \cite{bargerdeshpande}
{\renewcommand{\arraystretch}{1.3}
\begin{eqnarray}
\lefteqn{X =} \nonumber \\
 & & \frac{1}{2} \!\! \left( \begin{array}{ccc|c}
0 & 0 & 0 & 0 \\
0 & (g_1^2+g_2^2)(v_1^2+v_2^2) &
 g'\sqrt{g_1^2+g^2_2} (v_1^2 Y'_1 - v_2^2 Y'_2) &
 g''\sqrt{g_1^2+g^2_2} (v_1^2 Y''_1 - v_2^2 Y''_2) \\
0 & g'\sqrt{g_1^2+g^2_2} (v_1^2 Y'_1 - v_2^2 Y'_2) &
 {g'}^2 \sum_{i=1}^4 v_i^2 {Y'_i}^2 &
 g' g'' \sum_{i=1}^4 v_i^2 Y_i' Y_i''  \\ \cline{1-3}
0 & g''\sqrt{g_1^2+g^2_2} (v_1^2 Y''_1 - v_2^2 Y''_2) &
 \multicolumn{1}{c}{g' g'' \sum_{i=1}^4 v_i^2 Y_i' Y_i''} &
 {g''}^2 \sum_{i=1}^4 v_i^2 {Y''_i}^2
\end{array} \!\! \right) \!\! . \nonumber \\ \label{zmatrix}
\end{eqnarray}
}%
in the basis of the massless photon $A$ and the two [three] massive
gauge bosons $Z^\mathrm{SM}$, $Z'$ [and $Z''$] which mix.
$g_2$, $g_1$, $g'$ [and $g''$] denote the gauge couplings of the
SU(2)$_\mathrm{L}$, U(1)$_Y$, U(1)$'$ [and U(1)$''$] gauge
factors, respectively. 
The $Y'_i$ ($Y''_i$) are the U(1)$'$ (U(1)$''$) quantum numbers of the
doublet and singlet Higgs fields and the $v_i$ the respective vacuum
expectation values according to eq.~(\ref{higgsvev}).

In the rank-5 model the mixing matrix is obtained as the
upper left $3 \times 3$ submatrix of eq.~(\ref{zmatrix})
with the quantum numbers $Y'_i$ fixed by eq.~(\ref{Ysr5}).
With $\tan\beta = v_2/v_1$ and
\begin{equation}
  m_{Z^\mathrm{SM}}^2 \equiv \frac{m_W^2}{\cos^2\theta_W}
  = \frac{1}{2}(g_1^2+g_2^2)(v_1^2+v_2^2)
\end{equation}
the rank-5 mixing matrix becomes
\begin{equation}
  \addtolength{\arraycolsep}{3pt}\renewcommand{\arraystretch}{1.2}
X' = \left( \begin{array}{ccc} 0 & 0 & 0 \\
 0 & m_{Z^\mathrm{SM}}^2 & \delta m^2 \\
 0 & \delta m^2 & m_{Z'}^2
\end{array} \right)
\end{equation}
with
\begin{eqnarray}
 \delta m^2 & = & m_{Z^\mathrm{SM}}^2 \frac{g'}{g_1}
    \sin\theta_W \left( Y'_1 \cos^2\beta -
    Y'_2 \sin ^2\beta \right), \label{dm2}\\
 m_{Z'}^2   & = & \frac{1}{2} {Y'_3}^2 {g'}^2 v_3^2 + \frac{1}{2}
    {Y'_4}^2 {g'}^2 v_4^2 + m_{Z^\mathrm{SM}}^2
    \frac{{g'}^2}{g_1^2}\sin^2 \theta_W 
    \left( {Y'_1}^2 \cos^2\beta + {Y'_2}^2 \sin ^2\beta \right)
    .  \label{mzs2}
\end{eqnarray}
Thus the mass eigenstates $Z_1$ and $Z_2$ of the massive neutral gauge
bosons are
\begin{equation} \label{ZmassEZ}
\addtolength{\arraycolsep}{3pt}
\left( \begin{array}{c} Z_{1,\mu} \\ Z_{2,\mu} \end{array} \right)
= N^Z
\left( \begin{array}{c} Z^\mathrm{SM}_\mu \\ Z'_\mu \end{array} \right)
\end{equation}
with the orthogonal $2 \times 2$ diagonalization matrix
\begin{equation} \label{r5diag}
N^Z = \left( \begin{array}{rr} \cos\delta & \sin\delta \\
-\sin\delta & \cos\delta \end{array} \right),
\end{equation}
the mixing angle $\delta$,
\begin{equation} \label{deltadef}
\tan \delta = \frac{\delta m^2}{m_{Z^\mathrm{SM}}^2 - m_{Z_2}^2} \, ,
\end{equation}
and the mass eigenvalues
\begin{equation}
m^2_{Z_1,Z_2} = \frac{1}{2} \left( m_{Z^\mathrm{SM}}^2 + m_{Z'}^2 \mp
        \sqrt{(m_{Z^\mathrm{SM}}^2 - m_{Z'}^2)^2 + 4\delta m^4} \, \right) .
\end{equation}

In the rank-6 model with the full $4 \times 4$ mixing matrix
eq.~(\ref{zmatrix})
the quantum numbers of the doublet and singlet Higgs fields $Y'_i$ and
$Y''_i$  are fixed by eq.~(\ref{Ysr6}).
The submatrix of the massive gauge bosons $Z^\mathrm{SM}$, $Z'$ and
$Z''$ can be diagonalized by an orthogonal $3\times 3$ matrix $N^Z$
\begin{equation} \label{r6diag}
  \setlength{\arraycolsep}{4pt}
  \left( \begin{array}{cccc} 0 \\ & m_{Z_1} \\ & &  m_{Z_2} \\
      & & & m_{Z_3} \end{array} \right) =
  \left( \begin{array}{cccc} 1 & 0 & 0 & 0 \\ 0 \\ 0 &
      \multicolumn{3}{c}{N^Z} \\ 0 \end{array} \right) X
  \left( \begin{array}{cccc} 1 & 0 & 0 & 0 \\ 0 \\ 0 &
      \multicolumn{3}{c}{(N^Z)^\dag} \\ 0 \end{array} \right)
\end{equation}
with the mass eigenstates $Z_1$, $Z_2$ and $Z_3$.

\section{Chargino sector}
\setcounter{equation}{0}

The chargino mass term in the Lagrangian
\begin{equation}
  \mathcal{L}_{m_{\chi^\pm}} = -\frac{1}{2} (\psi^+ , \psi^-) \left(
    \begin{array}{cc} 0 & X^T \\ X & 0 \end{array}
  \right) {\psi^+ \choose \psi^-} + \textrm{h.c.}
\end{equation} \renewcommand{\arraystretch}{1.2}
with the chargino mass matrix
\begin{equation}
  X = \left( \begin{array}{cc} M_2 & \sqrt{2}\, m_W \sin\beta \\
      \sqrt{2}\, m_W \cos\beta & - \lambda v_3 \end{array} \right)
\end{equation}
is the same as in the MSSM if $-\lambda v_3$ is identified as
the MSSM parameter $\mu$  \cite{haberkane, gherghetta, bartlfraaschar}.

\section{Sfermion sector\label{sfermionmasses}}
\setcounter{equation}{0}

The mass terms of the sfermions can be derived from the scalar
potential \cite{boyce}. Because of the spontaneous breaking of the new
U(1) factors additional D-terms appear in comparison to the MSSM
\cite{cveticdemir, hr, drees,
boyce, keithma, gherghetta, MsfRGEsammel, Dtermsammel}.
Neglecting the mixing of the left and right sfermions which is small
for the first two generations the mass terms in the rank-5 [rank-6] model are
\begin{eqnarray}
 m_{\tilde{f}_L}^2 & = & \tilde{M}_{\tilde{f}_L}^2 + m_f^2 +
  L \, m_{Z^\mathrm{SM}}^2 \cos 2 \beta + Y'(f_L) \, \tilde{m}_{D'}^2
  \,[\!{}+ Y''(f_L) \, \tilde{m}_{D''}^2] \,, \label{msfermionr6_1} \\
 m_{\tilde{f}_R}^2 & = & \tilde{M}_{\tilde{f}_R}^2 + m_f^2
  - R \, m_{Z^\mathrm{SM}}^2 \cos 2 \beta - Y'(f_R) \, \tilde{m}_{D'}^2
  \,[\!{}- Y''(f_R) \, \tilde{m}_{D''}^2] \,. \label{msfermionr6_2}
\end{eqnarray}
$\tilde{M}_{\tilde{f}_{L,R}}$ are the scalar mass parameters and
\begin{equation} \label{LRdef}
  L = T_{3L} - Q \sin^2\theta_W \,, \quad R = - Q \sin^2\theta_W \,.
\end{equation}
In the rank-5 model with one singlet [two singlets]
the quantum numbers $Y'(f_{L,R})$ of the fermion fields are
\begin{equation} \label{Ysdef}
  Y'(f_L) = Y_\eta(f_L) \,, \qquad Y'(f_R) = Y_\eta(f_R) = -Y_\eta(f_L^c)
\end{equation}
as listed in Table~\ref{quantnr} and the
D-term is
\begin{equation}
  \tilde{m}_{D'}^2 = \frac{1}{4} {g'}^2
  \left( Y_1' v_1^2 + Y_2' v_2^2 + Y_3' v_3^2 \,[\!{}+ Y_4' v_4^2] \right)
\end{equation}
with $Y'_i$ according to eq.~(\ref{Ysr5}).

In the rank-6 model the quantum numbers are
\begin{equation} \label{YsYssdef}
  Y'(f_{L,R}) = Y_\psi(f_{L,R})\,, \qquad Y''(f_{L,R}) =
  Y_\chi(f_{L,R})
\end{equation}
as shown in Table~\ref{quantnr} and the two new D-terms read
\begin{eqnarray}
\tilde{m}_{D'}^2 & = & \frac{1}{4} {g'}^2
  \left( Y_1' v_1^2 + Y_2' v_2^2 + Y_3' v_3^2 + Y_4' v_4^2 \right) \,, \\
\tilde{m}_{D''}^2 & = & \frac{1}{4} {g''}^2
  \left( Y_1'' v_1^2 + Y_2'' v_2^2 + Y_3'' v_3^2 + Y_4'' v_4^2 \right)
\end{eqnarray}
with $Y'_i$, $Y''_i$ according to eq.~(\ref{Ysr6}).

\section{Lagrangians and couplings\label{lagrangians}}
\setcounter{equation}{0}

\subsection{Neutral currents Lagrangian}

In the rank-5 [rank-6] model the neutral currents Lagrangian reads
\cite{hr, london, LNCsammel}
\begin{eqnarray}
\mathcal{L}_\mathrm{NC} & = & {}- e Q \bar{f} \gamma^\mu f A_\mu
  \nonumber \\[1mm]
 & & {}- \frac{g_2}{\cos\theta_W} \bar{f}_L ( T_{3L} - Q
  \sin^2\theta_W ) \gamma^\mu f_L Z_\mu^{SM}
- \frac{g_2}{\cos\theta_W} \bar{f}_R ( - Q
  \sin^2\theta_W ) \gamma^\mu f_R Z_\mu^{SM} \nonumber \\[1mm]
 & & {}- g' \bar{f}_L \frac{Y'(f_L)}{2} \gamma^\mu f_L Z'_\mu
  - g' \bar{f}_R \frac{Y'(f_R)}{2} \gamma^\mu f_R Z'_\mu
  \nonumber \\[1mm]
 & & \!\!\left[\!{}- g'' \bar{f}_L \frac{Y''(f_L)}{2} \gamma^\mu f_L Z''_\mu
  - g'' \bar{f}_R \frac{Y''(f_R)}{2} \gamma^\mu f_R Z''_\mu\right]  .
   \label{LNCz1}
\end{eqnarray}
Here $e \equiv g_2 \sin\theta_W = \sqrt{4 \pi \alpha}$
is the absolute value of the electron charge and
$f$ denotes the respective fermion field with
$f_{L/R} = P_{L/R} f$ and the chiral projection operators
$P_{L,R} = \frac{1}{2} (1 \mp \gamma_5)$.
The $Y'(f_{L/R})$ and $Y''(f_{L/R})$ are the $\mathrm{U(1)'}$ and
$\mathrm{U(1)''}$ quantum numbers of the respective model as listed in
Table~\ref{quantnr}.

In terms of the mass eigenstates of the neutral gauge bosons
the Lagrangian has the form
\begin{equation}
  \mathcal{L}_\mathrm{NC} = - e Q \bar{f} \gamma^\mu f A_\mu
    - \sum_{n=1}^{n_Z} \frac{g_2}{\cos\theta_W} \bar{f} \gamma^\mu
      \left[L_n P_L + R_n P_R \right] f Z_{n,\mu}
\end{equation}
with eq.~(\ref{LRdef}) and
\begin{eqnarray}
  L_n & = & L N^Z_{n1} + \frac{Y'(f_L)}{2} \frac{g'}{g_1} \sin\theta_W N^Z_{n2}
    \,\Big[\!{}+ \frac{Y''(f_L)}{2} \frac{g''}{g_1} 
        \sin\theta_W N^Z_{n3}\Big] , \\[1mm]
  R_n & = & R N^Z_{n1} + \frac{Y'(f_R)}{2} \frac{g'}{g_1} \sin\theta_W N^Z_{n2}
    \,\Big[\!{}+ \frac{Y''(f_R)}{2} \frac{g''}{g_1} 
        \sin\theta_W N^Z_{n3}\Big] .
\end{eqnarray}
In the rank-5 model with $n_Z=2$ the couplings $Y'(f_{L,R})$ are given
in eq.~(\ref{Ysdef}) and $N^Z$ in eq.~(\ref{r5diag}), whereas in the
rank-6 model ($n_Z=3$) $Y'(f_{L,R})$, $Y''(f_{L,R})$ according to
eq.~(\ref{YsYssdef}) and $N^Z$ according to eq.~(\ref{r6diag}).

\subsection{\boldmath$Z$-neutralino-neutralino interaction}

The $Z$-neutralino-neutralino interaction Lagrangian in the rank-5
model with one singlet [with two singlets] has the form
\begin{eqnarray}
\mathcal{L}_{Z \tilde{\chi}^0 \tilde{\chi}^0}
& = & \frac{1}{4}
\left(g_2 W_\mu^3 - g_1B_\mu \right)
\left( \bar{\tilde{H_1}}\gamma^\mu \gamma_5 \tilde{H_1}-
\bar{\tilde{H_2}}\gamma^\mu \gamma_5 \tilde{H_2} \right)
\nonumber \\[1mm]
& & {} + \frac{1}{4}g'Z'_\mu \left(Y'_1
 \bar{\tilde{H_1}}\gamma^\mu \gamma_5 \tilde{H_1} + Y'_2
 \bar{\tilde{H_2}}\gamma^\mu \gamma_5 \tilde{H_2}
 + Y'_3 \bar{\tilde{N_1}}\gamma^\mu \gamma_5 \tilde{N_1}
 \,[\!{}+ Y'_4 \bar{\tilde{N_2}}\gamma^\mu \gamma_5 \tilde{N_2}]
\right),\qquad
\end{eqnarray}
whereas in the rank-6 model additional terms appear
\begin{equation}
{} + \frac{1}{4}g''Z''_\mu \left(Y''_1
 \bar{\tilde{H_1}}\gamma^\mu \gamma_5 \tilde{H_1} + Y''_2
 \bar{\tilde{H_2}}\gamma^\mu \gamma_5 \tilde{H_2}
 + Y''_3 \bar{\tilde{N_1}}\gamma^\mu \gamma_5 \tilde{N_1}
 + Y''_4 \bar{\tilde{N_2}}\gamma^\mu \gamma_5 \tilde{N_2}
\right).
\end{equation}
In both cases it can be written as
\begin{equation}
\mathcal{L}_{Z \tilde{\chi}^0 \tilde{\chi}^0} =
\sum_{n=1}^{n_Z}
\frac{1}{2} \frac{g_2}{\cos\theta_W} Z_{n,\mu} \bar{\tilde{\chi}^0_i}
\gamma^\mu
\left( O_{ij}^{''nL}P_L + O_{ij}^{''nR}P_R\right) \tilde{\chi}^0_j
\end{equation}
with 
\begin{eqnarray}
O_{ij}^{''nL}
& = & \frac{1}{2}
\left( \cos 2\beta (-N_{i3}N_{j3}^\ast + N_{i4}N_{j4}^\ast )
-\sin 2\beta (N_{i3}N_{j4}^\ast + N_{i4}N_{j3}^\ast )\right) N^Z_{n1}
\nonumber \\[1mm] & &
{}- \frac{1}{2} \frac{g'}{g_1} \sin\theta_W 
 \bigg( (Y'_1 \cos^2\beta + Y'_2 \sin^2\beta)
 N_{i3}N_{j3}^\ast + (Y'_1 \sin^2\beta + Y'_2 \cos^2\beta)
 N_{i4}N_{j4}^\ast  \nonumber \\ & & \hspace*{11mm}
{}+ \frac{1}{2} (Y'_1 - Y'_2) \sin 2\beta (N_{i3}N_{j4}^\ast +
 N_{i4}N_{j3}^\ast) + Y'_3 N_{i6}N_{j6}^\ast \,[\!{}+ Y'_4 N_{i7}N_{j7}^\ast]
 \bigg)  N^Z_{n2}  \nonumber \\[1mm] & &
\!\!\Big[\!{}- \frac{1}{2} \frac{g''}{g_1} \sin\theta_W 
 \bigg( (Y''_1 \cos^2\beta + Y''_2 \sin^2\beta) 
 N_{i3}N_{j3}^\ast + (Y''_1 \sin^2\beta + Y''_2 \cos^2\beta)
 N_{i4}N_{j4}^\ast  \nonumber \\[1mm] & & \hspace*{11mm}
{}+ \frac{1}{2} (Y''_1 - Y''_2) \sin 2\beta (N_{i3}N_{j4}^\ast +
 N_{i4}N_{j3}^\ast) + Y''_3 N_{i6}N_{j6}^\ast + Y''_4 N_{i7}N_{j7}^\ast
 \bigg)  N^Z_{n3}\Big] , \nonumber \\[-1mm] \\
O_{ij}^{''nR} & = & - \left( O_{ij}^{''nL}\right)^\ast,
\end{eqnarray}
where the diagonalization matrix $N$ of the neutralinos is given in
basis (\ref{r6basis}).
In the rank-5 models it is $n_Z = 2$ with the couplings $Y'_i$ according
to eq.~(\ref{Ysr5}) and the diagonalization matrix $N^Z$ of the neutral
gauge bosons according to eq.~(\ref{r5diag}). In the rank-6 model with
$n_Z = 3$ the couplings $Y'_i$ and $Y''_i$ are given in
eq.~(\ref{Ysr6}) and $N^Z$ in eq.~(\ref{r6diag}).

\subsection{Fermion-sfermion-neutralino interaction}

The fermion-sfermion-neutralino interaction Lagrangian has the same
form as in the MSSM
\begin{equation} \label{fsfneutww}
\mathcal{L}_{f\tilde{f}\tilde{\chi}^0_i} = g_2 f^L_{fi} \bar{f} P_R
\tilde{\chi}^0_i \tilde{f}_L + g_2 f^R_{fi} \bar{f} P_L
\tilde{\chi}^0_i \tilde{f}_R + \textrm{h.c.}
\end{equation}
with the extended couplings in the rank-5 [rank-6] model
\begin{eqnarray}
f^L_{fi} & = & - \sqrt{2} \Bigg( \frac{1}{\cos\theta_W} ( T_{3L} - Q
  \sin^2\theta_W ) N_{i2} + Q \sin\theta_W N_{i1}  \nonumber \\[1mm]
  & & \hspace*{11mm}{}+ \frac{Y'(f_L)}{2} \frac{g'}{g_1} \tan\theta_W N_{i5}
   \,\Big[\!{}+ \frac{Y''(f_L)}{2} \frac{g''}{g_1} \tan\theta_W
   N_{i8}\Big] \Bigg) \, , \\[2mm] 
f^R_{fi} & = & - \sqrt{2} \sin\theta_W \Bigg( Q \tan\theta_W N_{i2}^* -
  Q N_{i1}^* \nonumber \\
  & & \hspace*{23mm}{}- \frac{Y'(f_R)}{2} \frac{g'}{g_1}
   \frac{1}{\cos\theta_W} N_{i5}^* 
  \,\Big[\!{}- \frac{Y''(f_R)}{2} \frac{g''}{g_1}
   \frac{1}{\cos\theta_W} N_{i8}^*\Big] \Bigg) \, . 
\end{eqnarray}
$f$ denotes the respective fermion field, $\tilde{f}_{L/R}$ the
field of its scalar superpartner and $N$ the diagonalization matrix of
the neutralinos in basis (\ref{r6basis}).
In the rank-5 models the $Y'(f_{L,R})$ are given in eq.~(\ref{Ysdef}),
whereas in the rank-6 model $Y'(f_{L,R})$, $Y''(f_{L,R})$ are fixed
according to eq.~(\ref{YsYssdef}).

\end{appendix}

\end{document}